\documentclass[aps,superscriptaddress,showpacs,twocolumn,longbibliography,prb]{revtex4-2}

\usepackage{graphicx}
\usepackage{amssymb}
\usepackage{amsmath}
\usepackage{xspace}
\usepackage{color}
\usepackage{hyperref}
\usepackage{bm}
\usepackage{mathtools}

\begin{document}

\title{Cross-talk in superconducting qubit lattices with tunable couplers -- comparing transmon and fluxonium architectures}

\author{Florian Lange}
\affiliation{Friedrich-Alexander-Universit\"at Erlangen-N\"urnberg (FAU), Erlangen National High Performance Computing Center (NHR@FAU), 91058 Erlangen, Germany}
\author{Lukas Heunisch}
\affiliation{Physics Department, Friedrich-Alexander-Universit\"at Erlangen-N\"urnberg (FAU), 91058 Erlangen, Germany}
\author{Holger Fehske}
\affiliation{Friedrich-Alexander-Universit\"at Erlangen-N\"urnberg (FAU), Erlangen National High Performance Computing Center (NHR@FAU), 91058 Erlangen, Germany}
\affiliation{Institute of Physics, University of Greifswald, 17489 Greifswald, Germany}
\author{David P. DiVincenzo}
\affiliation{Institute for Quantum Information, RWTH Aachen University, 52056 Aachen, Germany}
\affiliation{J\"ulich-Aachen Research Alliance (JARA), Fundamentals of Future Information Technologies, 52425 J\"ulich, Germany}
\affiliation{Peter Gr\"unberg Institute, Theoretical Nanoelectronics, Forschungszentrum J\"ulich, 52425 J\"ulich, Germany}
\author{Michael J. Hartmann}
\affiliation{Physics Department, Friedrich-Alexander-Universit\"at Erlangen-N\"urnberg (FAU), 91058 Erlangen, Germany}

\date{\today}

\begin{abstract}
	Cross-talk between qubits is one of the main challenges for scaling superconducting quantum processors. Here, we use the density-matrix renormalization group to numerically analyze lattices of superconducting qubits from a perspective of many-body localization. 
	Specifically, we compare different architectures that include tunable couplers designed to decouple qubits in the idle state, and calculate the residual $ZZ$ interactions as well as the inverse participation ratio in the computational basis states. 
	For transmon qubits outside of the straddling regime, the results confirm that tunable C-shunt flux couplers are significantly more efficient in mitigating the $ZZ$ interactions than tunable transmons. 
	A recently proposed fluxonium architecture with tunable transmon couplers 
	is demonstrated to also maintain its strong suppression of the $ZZ$ interactions in larger systems, while having a higher inverse participation ratio in the computational basis states than lattices of transmon qubits. Our results thus suggest that fluxonium architectures may feature lower cross talk than transmon lattices when designed to achieve similar gate speeds and fidelities. 
\end{abstract}

\maketitle

\section{Introduction}

Superconducting qubits, in particular transmons~\cite{TransmonQubit}, are currently the key components of leading quantum-computing architectures~\cite{SuperconductingQubitGuide,SuperconductingQubitsStateOfPlay}. 
A common problem in such devices is that the couplings between qubits, which enable the application of entangling gates, also lead to unwanted static $ZZ$ interactions that cause dephasing of the qubits and are known to hinder the implementation of gates with high fidelity~\cite{PhysRevApplied.14.024042,PhysRevLett.125.200504}.

Typical superconducting qubit chips have a nearest-neighbor connectivity on a two-dimensional grid~\cite{SuperconductingQubitsStateOfPlay,QCNearTermChallenges}. 
In the idle mode, such a system can be described approximately by a time-independent lattice Hamiltonian with short-ranged couplings.  
Variations in the qubit properties, either engineered or due to imprecisions in the fabrication, show up as disorder in the Hamiltonian. 
This is important, as it causes the system to be in a many-body-localized 
phase with emergent local integrals of motion, so called $l$-bits, that can be identified as dressed qubit degrees of freedom~\cite{TransmonsChallengedByChaos}. 
Important questions regarding the operability of such a device are then (i) how strongly the effective qubits are localized and (ii) how large the residual couplings between them are. 
The latter includes the static $ZZ$ interaction but also its generalization to terms involving three or more qubits, which can become dominant in certain parameter regions~\cite{PhysRevApplied.22.064030}. 

Previous studies of fixed-frequency transmons have shown that the device parameters need to be chosen carefully to 
avoid approaching a quantum-chaotic regime as the number of qubits is increased, which would delocalize the qubits and lead to the emergence of problematic longer-ranged couplings~\cite{TransmonsChallengedByChaos}. 
Alternative architectures that employ tunable couplers as additional circuit elements are expected to be more stable in this regard, since they can approximately decouple pairs of neighboring qubits in the idle state. 
However, a residual $ZZ$ interaction generally remains when flux-tunable transmons are used as couplers outside of the straddling regime~\cite{PhysRevLett.125.240502,PhysRevLett.127.080505}. 
More elaborate schemes may enable a full decoupling at least for individual qubit pairs~\cite{DoubleTransmonCoupler,CShuntFluxCoupler}.  
One such proposal is to employ C-shunt flux couplers, which differ from transmons by their positive anharmonicity~\cite{PhysRevLett.125.200503,PhysRevLett.125.200504,CShuntFluxCoupler}.

In this work, we consider a 2-leg ladder of qubits and numerically study the localization properties and the static interactions between qubits for both transmon and C-shunt flux couplers. 
We further do simulations for fluxonium qubits~\cite{FluxoniumWithTransmonCoupler}, which are promising candidates for the next qubit generation due to their large anharmonicity and the improvement in coherence times, focusing on a recently proposed scheme with transmon couplers that was shown to exhibit a strong suppression of $ZZ$ interactions in two-qubit systems~\cite{FluxoniumWithTransmonCoupler}. 
Our calculations are based on an extension of the density-matrix renormalization group~\cite{DMRG} (DMRG) to many-body-localized systems in one dimension~\cite{DMRG-X}. 
By using matrix-product-state (MPS) approximations for the computational basis states, it is possible to efficiently simulate systems of $8$ qubits and $10$ couplers, 
which provides access to static interactions beyond the nearest-neighbor $ZZ$ couplings and gives an indication of 
how well different architectures scale to a larger number of qubits.

The rest of the paper is organized as follows: Section~\ref{sec:method} introduces the model for the superconducting qubit systems and the theoretical approach. In Sec.~\ref{sec:transmons}, we analyze transmon systems coupled in three different ways: directly via capacitive coupling, via an additional tunable transmon, and using a C-shunt flux coupler, while Sec.~\ref{sec:fluxonium} deals with fluxonium systems coupled via transmon tunable couplers. 
We end with a discussion of the results and a summary in Secs.~\ref{sec:discussion} and~\ref{sec:summary}, respectively.

\section{Model and methodology}
\label{sec:method}

\subsection{System Hamiltonian}
\begin{figure}[t]
  \centering
  \includegraphics[width=0.85\linewidth]{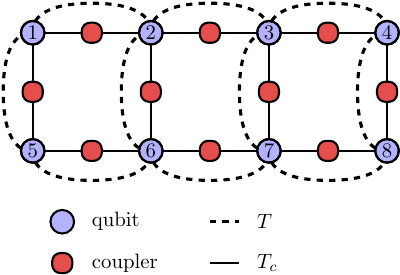}
  \caption{Layout of the considered systems of superconducting qubits and couplers.}
  \label{fig:sys2x4}
\end{figure}

We consider models of capacitively coupled superconducting qubits and tunable couplers that are arranged as 
a two-leg ladder (see Fig.~\ref{fig:sys2x4}). Their Hamiltonians are of the form 
\begin{align}
	\hat{H} &= \hat{H}_{\text{loc}} + \sum_{\mathclap{\langle q_1 q_2 c \rangle}} \left[ T \hat{n}_{q_1} \hat{n}_{q_2} + T_c (\hat{n}_{q_1} \hat{n}_c + s \, \hat{n}_{q_2} \hat{n}_c) \right]  \, ,
	\label{eq:qubitlattice}
\end{align}
where $\hat{H}_{\text{loc}}$ is the sum of the local terms, $T$ and $T_c$ are the coupling strengths between the qubits and between the qubits and the coupler, respectively, $\hat{n}_i$ is the Cooper-pair number operator for the qubit or coupler with index $i$, and $\langle q_1 q_2 c \rangle$ denotes a pair of qubits and the corresponding coupler $c$. 
Equation~\eqref{eq:qubitlattice} should be regarded as an approximation of a circuit with coupling capacitances between the qubits and couplers connected by lines in Fig.~\ref{fig:sys2x4}. 
A more accurate treatment would involve extracting the couplings from the inverse of the capacitance matrix, including longer-ranged terms that we neglect here, since they are typically at least an order of magnitude smaller than the couplings considered in Eq.~\eqref{eq:qubitlattice}. 

Depending on its architecture, the tunable coupler can couple to its neighboring qubits either with the same sign $(s=1)$ or with opposite signs $(s=-1)$. While floating architectures with $s=-1$ are preferred for C-shunt flux couplers (see ~\cite{CShuntFluxCoupler}), we also analyze floating transmon coupler architectures that couple to qubits with opposite signs. This allows for a more direct comparison between both architectures. For the fluxonium system, we adopt the architecture described in ~\cite{FluxoniumWithTransmonCoupler}, where the transmon couplers couple to fluxonium qubits with the same sign. 
The local terms $\hat{H}_{\text{loc}}$ depend on the type of qubit or coupler and are discussed in the corresponding sections. They can also include disorder, which leads to many-body localization (MBL) of the system~\cite{MBLAnnualReview}.

\subsection{MBL states in superconducting architectures}
The Hamiltonian of a spin-1/2 system in a fully many-body localized phase can be written as~\cite{MBLAnnualReview,PhysRevLett.111.127201,MBLPhenomenology,MBLColloquium}
\begin{align}
	\hat{H}_{\text{MBL}} = \sum_{\bm{n} \in \{0,1\}^N} w_{\bm{n}} \bigotimes_{i=1}^N \hat{\tau}_i^{n_i} \, ,
  \label{eq:l-bits}
\end{align}
where $N$ is the number of sites, $w_{\bm{n}}$ are real coefficients and $l$-bits $\hat{\tau}_i$ are 
operators related to the Pauli $Z$ operators by a quasi-local unitary transformation. 
Bitstrings $\bm{n}$ of length $N$ are used to label the different terms in the operator expansion, with $n_i = 1$ indicating a $\hat{\tau}$ operator at site $i$. 
The definition of the $l$-bit operators is not unique. One way to construct a set of $l$-bits is to diagonalize the Hamiltonian, define an injective map from the product states $|\bm{n}\rangle = \otimes_i |n_i\rangle $ to the eigenstates $| \tilde{\bm{n}}\rangle$ that maximizes the sum of the overlaps $\sum_{\bm{n} \in \{0,1\}^N} |\langle \bm{n} | \tilde{\bm{n}} \rangle|^2$~\cite{MBLEncodingWithMPO}, and set $\hat{\tau}_i = \sum_{\bm{n}} (-1)^{n_i} | \tilde{\bm{n}} \rangle \langle \tilde{\bm{n}} |$. 
Alternatively, one may use a mapping based on flow equations~\cite{PhysRevLett.119.075701} or tracing the energy levels starting from a decoupled limit with maximum localization~\cite{TransmonsChallengedByChaos}. 

Models of superconducting qubits like Eq.~\eqref{eq:qubitlattice} also fit into the above framework, 
with $|\tilde{\bm{n}}\rangle$ becoming the computational basis states and $\hat{\tau}_i$ the $Z$ operators for the dressed qubits. 
A difference compared with the spin-1/2 case is that the systems include higher-lying excited states for the qubits, and tunable couplers as additional sites. 
The computational basis states are therefore only those eigenstates that are matched with the product states in which the qubits are in one of the two lowest energy levels and the couplers are in their ground states (ignoring coupling terms between qubits or couplers). 
After projecting onto the $2^N$-dimensional computational subspace, the Hamiltonian again has the form~\eqref{eq:l-bits}, 
which defines the longitudinal interactions between idling qubits. 
The coefficients $w_{\bm{n}}$ describe how the frequency of a qubit shifts depending on the state of the other qubits and could thus be measured in experiments.

In addition to these interactions, we want to investigate
how well the computational basis is localized, i.e., how much the dressed basis states $|\tilde{\bm{n}}\rangle$ differ from the corresponding product states $|\bm{n}\rangle$. 
A possible measure for this is the inverse participatio ratio (IPR). 
For a state $|\psi\rangle$, it is defined as
\begin{align}
	\text{IPR} &= \sum_{\bm{n}} | \langle \psi | \bm{n} \rangle |^4 \, ,
\end{align}
where the sum is over all states in the undressed basis. 
As the interaction coefficients, the IPR can be probed in experiments as well. By applying a Rabi pulse to one qubit and testing whether a neighboring qubit has also been flipped, one could measure how well the qubit states are localized.

To determine the coupling coefficients $w_{\bm{n}}$ and the IPR numerically, 
we make use of the fact that MBL states in one dimension admit an efficient approximation as MPS~\cite{SCHOLLWOCK201196} and employ the DMRG-X method~\cite{DMRG-X}. 
The difference compared to the regular DMRG method is that it targets eigenstates based on their overlap with a reference state, in this case the relevant product states $|\bm{n}\rangle$. 
Unless noted otherwise, the mapping between product states and eigenstates simplifies to picking the eigenstate with the largest overlap, and one DMRG-X calculation for each $n$ is sufficient. If two product states map to the same eigenstate, we do additional DMRG-X simulations with an orthogonalization against the previously calculated states, and switch to the more general mapping mentioned above that maximizes the sum of overlaps~\cite{MBLEncodingWithMPO}. 

The interaction coefficients are obtained 
from the energies $E_{\bm{n}}$ by a 
Walsh-Hadamard transformation~\cite{TransmonsChallengedByChaos}
\begin{align}
	w_{\bm{n}} &= \frac{1}{2^N} \sum_{\bm{n}'\in \{0,1 \}^N} (-1)^{\sum_i n_i n_i'} E_{\bm{n}'} \, .
\end{align}
Calculating the IPR from an MPS representation is also straightforward, though the computational cost scales as $\chi^5$ in the bond dimension $\chi$
when using a direct evaluation. 
For large bond dimensions, it is therefore more efficient to employ a stochastic method 
scaling like $\chi^3$, where one samples from the MPS in the undressed basis. 

Because of the larger local Hilbert spaces, we used a single-site variant of the DMRG-X algorithm. 
As long as the bond dimension $\chi$ is small, the local eigenproblems in the DMRG-X can be efficiently solved by applying a standard solver 
to the dense-matrix representation of the projected Hamiltonians. 
For larger $\chi$, we switched to a matrix-free Krylov method and look for the eigenvector to the smallest eigenvalue of the $(\hat{H} - \lambda)^2$,  
where $\lambda$ is the approximate energy from a calculation with small $\chi$. 
We checked the convergence for each state by monitoring the variance $\langle \tilde{\bm{n}} | \hat{H}^2 | \tilde{\bm{n}} \rangle - \langle \tilde{\bm{n}} | \hat{H} | \tilde{\bm{n}} \rangle^2$. 
In cases where the calculations did not converge to the targeted eigenstate with sufficient accuracy straightaway, usually because the DMRG simulations got stuck in a superposition of two eigenstates close in energy, we found it helpful to apply an iteration of the shift-invert DMRG method~\cite{ShiftInvertDMRG}, implemented with the library Krylov.jl~\cite{KrylovJL}. 
To expand the MPS bonds, we used the method described in Ref.~\cite{MPS_AMEn}. 

We note that a DMRG-X approach was used recently to study lattices of superconducting qubits. In that work, a multi-target version of the algorithm was proposed to improve convergence for strongly hybridized states~\cite{MTDMRGX}. 

\section{Transmon qubits}
\label{sec:transmons}
The Hamiltonian of a single transmon with charging energy $E_C$ and Josephson energy $E_J$ is~\cite{ciani2024lecture}
\begin{align}
  \hat{H}_{\text{T}} &= 4 E_C \hat{n}^2 - E_{J} \cos(\hat{\phi}) \, ,
\end{align}
where $\hat{\phi}$ is the superconducting phase that obeys the commutation relation $[\hat{\phi},\hat{n}] = \text{i}$ with the Cooper-pair number operator $\hat{n}$.
It is convenient to also introduce the ladder operators for the quadratic part of $\hat{H}_\text{T}$ via 
\begin{align}
  \hat{\phi} &= \left( \frac{2 E_C}{E_{J}} \right)^{\frac{1}{4}} (\hat{b} + \hat{b}^\dagger) \, , \\
  \hat{n} &= \text{i}\left( \frac{E_{J}}{32 E_C} \right)^{\frac{1}{4}} (\hat{b} - \hat{b}^\dagger) \, .
  \label{eq:cooperpairop_transmons}
\end{align}
Considering the transmon regime, where $E_J \gg E_C$, the cosine term can be expanded in powers of $\phi$, which leads to the widely accepted modeling of transmons as Duffing oscillators: 
\begin{align}
  \hat{H}_{\text{T}} &\approx \nu \, \hat{b}^\dagger \hat{b} + \frac{\alpha}{2} \hat{b}^\dagger \hat{b}^\dagger \hat{b} \hat{b} \, ,
  \label{eq:duffing}
\end{align}
with $\nu = \sqrt{8 E_C E_J} - E_C$ and $\alpha = -E_C$. In the following we truncate the Hilbert space to the lowest $6$ eigenstates and study lattices of superconducting qubits described by the Hamiltonian~\eqref{eq:qubitlattice} with local terms as in Eq.~\eqref{eq:duffing}. The charging energy is assumed to be $E_C = 0.2 \, \text{GHz}$ for all qubits, while the Josephson energy $E_J$ is site-dependent, introducing disorder in the qubit frequencies $\nu$. 
Throughout this work, we set $h=1$.

Clearly, the choice of disorder will strongly affect the localization of the qubits and their residual interactions. A common scheme is to have different mean frequencies between sublattices to reduce the hybridization between neighboring qubits. In addition, quasi-periodic distributions have recently been proposed to prevent resonances also between qubits that are further apart~\cite{WalshQuasiPeriodic}. 
We follow Ref.~\cite{WalshQuasiPeriodic} and use a quasi-periodic Aubry-Andr\'{e} model for the qubit frequencies~\cite{AubryAndreModel}:
\begin{align}
  \nu_i &= \bar{\nu} + \Delta \sqrt{2} \sin \Big[ \pi x_i \big( y_i + \sqrt{y_i^2 + 4} \big) \Big] \, .
\end{align}
In an infinite ladder, $\bar{\nu}$ and $\Delta$ become the average frequency and the standard deviation, respectively. 
We choose $\bar{\nu} = 5.8\,\text{GHz}$, $\Delta = 0.33\,\text{GHz}$, and an $8$-qubit lattice with $x_i \in \{1,2,3,4\}$ and $y_i \in \{ 1, 2 \}$ (see Table~\ref{table:transmonparameters}). 
For the couplings between transmon qubits, we consider three different cases: the ladder (i) with only static capacitive coupling, (ii) with tunable transmons and (iii) with tunable C-shunt flux couplers.

\begin{table}[tb]
  \centering
        \newcommand{\taborder}[4]{#1 & #2 & #3 & #4}
  \setlength{\tabcolsep}{8pt}
  \begin{tabular}{c|ccc}
	  \taborder{qubit}{$E_C$}{$E_J$}{$\nu$}\\
          \hline
          \taborder{$1$}{ 0.2}{24.33}{6.040} \\
          \taborder{$2$}{ 0.2}{19.52}{5.389} \\
          \taborder{$3$}{ 0.2}{26.13}{6.266} \\
	  \taborder{$4$}{ 0.2}{19.68}{5.411} \\
	  \taborder{$5$}{ 0.2}{20.20}{5.485} \\
          \taborder{$6$}{ 0.2}{26.12}{6.265} \\
	  \taborder{$7$}{ 0.2}{19.81}{5.430} \\
	  \taborder{$8$}{ 0.2}{23.11}{5.881} \\[1mm]
  \end{tabular}
	\caption{Parameters of the transmon system with the layout shown in Fig.~\ref{fig:sys2x4}. All values are in GHz.}
  \label{table:transmonparameters}
\end{table}

\subsection{Static capacitive coupling}
\begin{figure}[tb]
  \centering
  \includegraphics[width=0.97\linewidth]{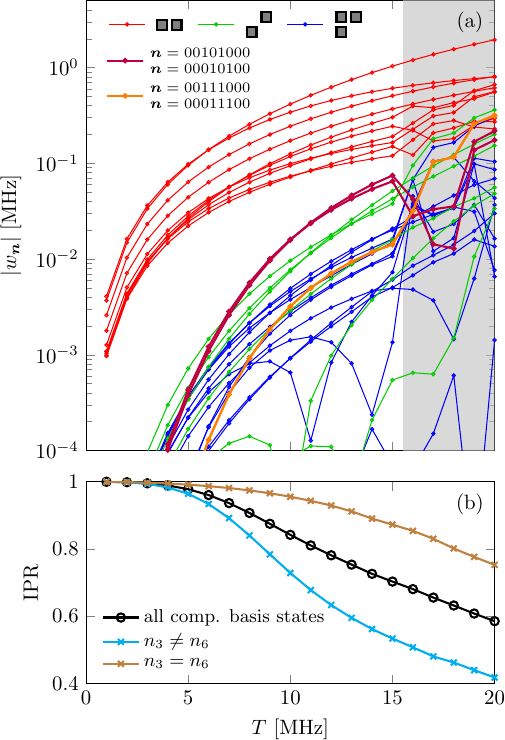}
	\caption{(a) Interaction coefficients $w_{\bm{n}}$ for transmons without couplers on a ladder. The graphical labels indicate the positions of the involved $Z$ operators (up to rotations and translations), e.g., the red curves show horizontal and vertical nearest-neighbor couplings, and the green ones the next-nearest-neighbor couplings along the diagonals. Also included are the specific longer-ranged couplings discussed in the main text (purple and orange curves). In the shaded region, the state assignment is ambiguous. The dips in some curves signal sign changes of the corresponding interaction terms. (b) Average IPR for all computational basis states and separated based on the states of the qubits $3$ and $6$.
        }
  \label{fig:walsh_nocouplers}
\end{figure}

Figure~\ref{fig:walsh_nocouplers}(a) displays the interaction coefficients $w_{\bm{n}}$ as a function of the capacitive-coupling strength $T$ 
for a system without tunable couplers ($T_c=0$). 
Naturally, the dominant $ZZ$ couplings are those between nearest neighbors, which 
reach values on the order of $10\,\text{kHz}$ already around $T\approx 2\,\text{MHz}$. 
Longer-ranged terms are orders of magnitude smaller at weak capacitive coupling, but become comparable in strength as $T$ increases, indicating that the delocalized regime is approached. 
In addition to stronger interactions, increasing $T$ leads to a delocalization of the dressed qubits. 
This is captured by the average IPR of the computational basis states, which remains close to $1$ up to around $T=5\,$MHz before it drops off and reaches values of approximately $0.6$ for $T=20\,$MHz [Fig.\ref{fig:walsh_nocouplers}(b)].
Overall, our results confirm previous exact-diagonalization studies of MBL in transmon arrays~\cite{TransmonsChallengedByChaos,WalshQuasiPeriodic,ClassicalChaosTransmons}.

It is expected that resonant clusters of sites form as an MBL system gets closer to the delocalization transition~\cite{PhysRevX.5.031033}.
Here, the first sign of this appears in interaction coefficient for 
the next-nearest neighbor qubits 3 and 6, whose frequencies $\nu_i$ are almost degenerate. 
An indication of the strong hybridization is that
the IPR is much lower for states with a single excitation between the two qubits.  
We also observe substantial long-ranged couplings $w_{\bm{n}}$ involving qubits 3 and 6 
that correspond to the bitstrings $\bm{n} \in \{ 00101000, 00010100, 00111000, 00011100 \}$ in Fig.~\ref{fig:walsh_nocouplers}(a). 
For $T > 15$\,MHz, the definition of the computational basis states becomes ambiguous, as picking the eigenstate with the largest overlap is no longer an injective map. 
The results shown 
are based on the state assigment described in Sec.~\ref{sec:method}. 
There is still an overall increase of the interaction strengths $|w_{\bm{n}}|$ with $T$, but with jumps for some coefficients due to changes in the state mapping. 
One could also assign the computational basis states by tracing the spectrum as a function of $T$ and identifying the anticrossings where the state labels switch~\cite{TransmonsChallengedByChaos}. Since that method is based on the energy, it might lead to a less erratic $T$-dependence of the interaction coefficients for large capacitive couplings. 
It is clear, however, that the system in this regime would not be suitable as a quantum-computing device and that any method of state assignment breaks down for sufficiently large $T$ when the system becomes delocalized.

\subsection{Tunable transmon couplers}
Let us now add flux-tunable transmons between each pair of directly coupled qubits. These couplers can be approximately described by the same Hamiltonian~\eqref{eq:duffing} as the fixed-frequency transmons. An important consideration is the choice of the coupler parameters,
which ideally should enable fast gate operations with high fidelity, while decoupling the qubits as much as possible in the idle regime. 
The key idea is that a tunable coupler introduces a second coupling channel that can interfere destructively with the direct capacitive coupling. However, full cancellation of interactions between qubits is only achievable within the straddling regime for transmon tunable couplers, where the detuning between a pair of coupled qubits is smaller than the absolute value of the anharmonicity $\alpha$~\cite{PhysRevLett.127.080505}. Outside this regime, residual $ZZ$-type interactions remain present, limiting the system’s performance. On the other hand, constraining the system to this parameter regime enhances problems such as frequency crowding, making a larger detuning between qubits more desirable.

\begin{figure}[tb]
  \centering
  \includegraphics[width=0.97\linewidth]{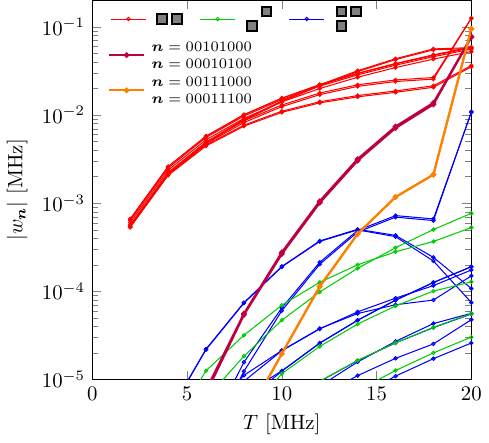}
	\caption{Strengths of the residual interactions in a system with transmon qubits and tunable transmon couplers. The notation is the same as in Fig.~\ref{fig:walsh_nocouplers}(a).}
  \label{fig:walsh_transmon_coupler}
\end{figure}

The model we consider is outside the straddling regime, where a residual $ZZ$ coupling remains, but a strong reduction compared to an architecture without couplers is possible. 
To find an appropriate idle point for the DMRG analysis of the system, we set $E_{C}=0.15$\,GHz for each coupler and determine the Josephson energy $E_{J}$ that minimizes the $ZZ$ interaction of the corresponding isolated qubit pair. For simplicity, we also fix $T_c^2/T = 1.2 \, \text{GHz}$, so that the idle points of the couplers do not change notably as $T$ is varied.

As shown in Fig.~\ref{fig:walsh_transmon_coupler}, the strength of the nearest-neighbor terms is reduced by about an order of magnitude compared to the system without tunable couplers, and there is a clearer separation in energy scales to longer-ranged interactions. 
Specific long-ranged terms related to the qubits 3 and 6 still become significant for large enough $T$, however, indicating that these tunable couplers are not able to fully remove the harmful effect of nearly resonant qubits. 
At very strong couplings $T=20\,$MHz, there are jumps in some of the interaction coefficients $w_{\bm{n}}$, 
 likely caused by the switching of state labels at avoided level crossings~\cite{TransmonsChallengedByChaos}. 
  Such difficulties in assigning the computational basis states are pushed to larger $T$ compared to the system without tunable couplers.

\begin{figure}[tb]
  \centering
  \includegraphics[width=0.9\linewidth]{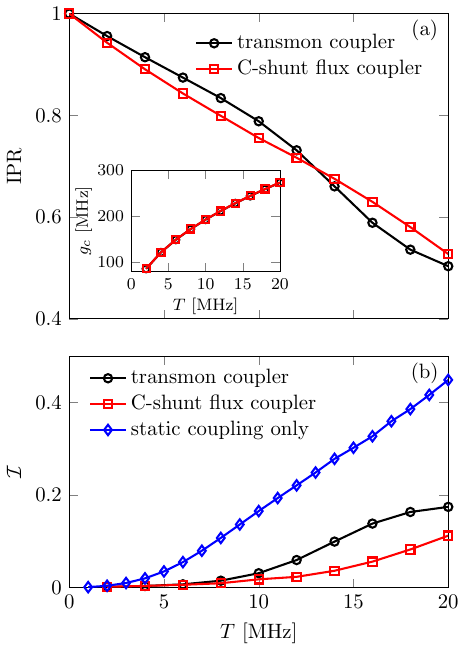}
	\caption{(a) IPR averaged over all states of the computational basis for a ladder with either tunable transmons or C-shunt flux qubits as couplers. The inset shows the average hopping amplitudes between qubits and couplers, taking the factors in Eqs.~\eqref{eq:cooperpairop_transmons} and ~\eqref{eq:cooperpairop_cshunt} into account. (b) average mutual information between a qubit and the remaining system excluding the connected couplers. 
        } 
  \label{fig:IPR}
\end{figure}

The IPR is significantly lower compared to the system with static capacitive coupling (see Fig.~\ref{fig:IPR}), especially for small coupling constants $T$. This can be attributed to the fact that, although the qubits appear decoupled in a Schrieffer-Wolff picture~\cite{SchriefferWolffReview}, their states experience significant dressing due to strong couplings with neighboring couplers. 
To distinguish this hybridization from problematic longer-ranged delocalization, 
we examine the quantum mutual information between the qubit and the rest of the system, excluding the directly coupled tunable couplers~[Fig.~\ref{fig:IPR}(b)].
Labeling the subsystems $A$ and $B$, respectively, the mutual information for a state $|\psi\rangle$ is
\begin{align}
	\mathcal{I}(A,B) &= S(\rho_A) + S(\rho_B) - S(\rho_{A \cup B}) \, ,
\end{align}
with $S(\rho) = -\text{Tr}(\rho \ln \rho)$ the von Neumann entanglement entropy and $\rho_X = \text{Tr}_{\bar{X}}( |\psi \rangle \langle \psi |)$ the reduced density matrix of subsystem $X$ (with complement $\bar{X}$). 
This quantity, averaged over all computational basis states and qubits, does not increase much until around $T=10\,$MHz, whereas the corresponding quantity in the model without couplers already starts to rise around $T=5\,$MHz. 
Tunable couplers thus appear to be effective in keeping the qubits localized in the sense that they reduce the spread of the actual (slightly hybridized) computational basis to other qubit sites.

\subsection{C-shunt flux couplers}
A promising approach to further suppress qubit crosstalk is the use of a hybrid architecture that combines two qubit types with opposite-sign anharmonicities. While Ref.~\cite{PhysRevLett.125.200504} proposes reducing crosstalk by coupling transmons with negative anharmonicities to flux qubits with a positive anharmonicity via a tunable bus, Ref.~\cite{CShuntFluxCoupler} suggests that the positive anharmonicity of C-shunt flux qubits can also be leveraged in a coupling device. This allows to couple well-established transmon architectures in a scalable manner without requiring integration of multiple qubit types on a single chip. 
A single C-shunt flux coupler is described by the Hamiltonian
\begin{align}
	\hat{H}_{\text{CSF}} &= 4 E_C \hat{n}^2 - 2 E_J \cos \left( \frac{\hat{\phi}}{\sqrt{2}} \right) \nonumber \\ 
	&\hspace*{5mm} - \gamma E_J \cos\left(\sqrt{2} \hat{\phi} + \frac{\phi_{\text{ext}}}{2\pi} \right) \, ,
\end{align}
where $\gamma \in \mathbb{R}^+$ and $\phi_{\text{ext}} = 2\pi \Phi_\text{ext} / \Phi_0$ is the reduced flux threading the loop of three Josephson junctions. 
The parameters $E_J$, $\gamma$ and $\phi_{\text{ext}}$ should be fixed appropriately
to decouple the qubits as much as possible in the idle state. 
One may also use $\phi_{\text{ext}}$ as a control parameter to implement two-qubit gates.

\begin{figure}[tb]
  \centering
  \includegraphics[width=0.97\linewidth]{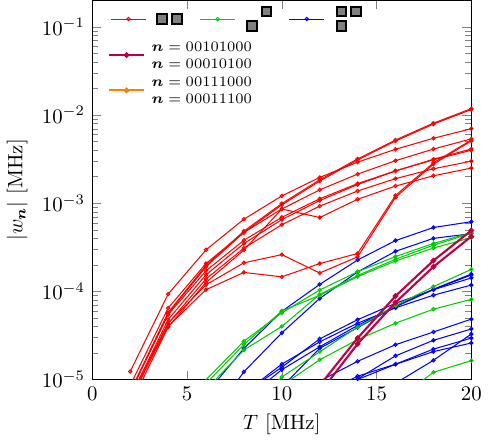}
	\caption{Same as Fig.~\ref{fig:walsh_transmon_coupler} but for tunable C-shunt flux couplers.
        } 
  \label{fig:walsh_cshunt_coupler}
\end{figure}

An approximate model similar to Eq.~\eqref{eq:duffing} 
for transmons is obtained by calculating the minimum $\phi_{\text{min}}$ of the potential for the phase variable, which satisfies $\sin(\phi_\text{min}/\sqrt{2}) + \gamma \sin(\sqrt{2} \phi_\text{min} + \phi_\text{ext})=0$, and then expanding around it. Expressing the Cooper pair number $\hat{n}$ and flux $\hat{\Psi}$ in terms of bosonic creation and annihilation operators $\hat{b}$ and $\hat{b}^\dagger$, one obtains:
\begin{align}
	\hat{H}_{\text{CSF}} 
	&\approx \nu \,\hat{b}^\dagger \hat{b} + \frac{\alpha}{2} \hat{b}^\dagger \hat{b}^\dagger \hat{b} \hat{b} \nonumber \\
	& \hspace*{5mm} + 3 K_3 \left( \frac{E_C}{K_2} \right)^{\mathclap{\frac{3}{2}}} ( \hat{b}^\dagger + \hat{b} + \hat{b}^\dagger \hat{b}^\dagger \hat{b} +  \hat{b}^\dagger \hat{b} \hat{b} )  \, ,
\end{align}
with
\begin{align}
        K_2 &= \frac{E_J}{2} \bigg[ \cos\left(\frac{\phi_\text{min}}{\sqrt{2}}\right) \nonumber \\ & \hspace*{1.0cm} + 2 \gamma \cos \big( \sqrt{2} \phi_\text{min} + \phi_\text{ext}\big) \bigg] \, , \\
        K_3 &= -\frac{E_J}{6} \bigg[ \frac{1}{\sqrt{2}} \sin\left(\frac{\phi_\text{min}}{\sqrt{2}}\right) \nonumber \\ & \hspace*{1.0cm} + 2 \sqrt{2} \gamma \sin \big( \sqrt{2} \phi_\text{min} + \phi_\text{ext}\big) \bigg] \, , \\
        K_4 &= -\frac{E_J}{24} \bigg[ \frac{1}{2} \cos\left(\frac{\phi_\text{min}}{\sqrt{2}}\right) \nonumber \\ & \hspace*{1cm} + 4 \gamma \cos \big( \sqrt{2} \phi_\text{min} + \phi_\text{ext}\big) \bigg] \, .
\end{align}
The qubit frequency and anharmonicity are
\begin{align}
	\nu &= 4 \sqrt{E_C K_2} + \alpha \, , \\
	\alpha &= 12 \frac{E_C K_4}{K_2} \, ,
\end{align}
and the boson operators are defined by
\begin{align}
  \hat{\phi} &= \left(\frac{E_C}{K_2} \right)^{\frac{1}{4}} (\hat{b} + \hat{b}^\dagger) \, , \\
  \hat{n} &= \text{i}\left( \frac{K_2}{16 E_C} \right)^{\frac{1}{4}} (\hat{b} - \hat{b}^\dagger) \, .
  \label{eq:cooperpairop_cshunt}
\end{align}
We have truncated the Hilbert space to a maximum of $5$ bosons in our simulations. 

Previous work has demonstrated that the nearest-neighbor $ZZ$ coupling can be fully removed even when spectator qubits are included, although the ideal parameter points may change noticeably compared with isolated qubit pairs~\cite{CShuntFluxCoupler}. In order to have a fair comparison with transmon couplers, we first pick the parameters based on isolated qubit dimers as before. We again set $E_C=0.15$\,GHz for the couplers and keep the flux parameter fixed at $\phi_{\text{ext}} = 0.51$, which is slightly detuned from the point of maximum anharmonicity $\alpha$ and minimum frequency $\nu$ to allow for a fine tuning of $\nu$ in both directions at a later stage. The condition of vanishing $ZZ$ coupling defines a curve in the $\gamma$-$E_J$ plane. We minimize the parameter
\begin{align}
  \epsilon &= \text{max}\big( |\langle \widetilde{100} | 001 \rangle |^2, \; |\langle \widetilde{001} | 100  \rangle |^2 \big)
  \label{eq:delocalization_parameter}
\end{align}
along this curve to additionally keep the dressed qubit states as localized as possible. Here, $|\widetilde{q_1 c \, q_2} \rangle$ and $|q_1 c \, q_2 \rangle$ label, respectively, the dressed and undressed states of the qubits $q_{1,2}$ and coupler $c$.

Figure.~\ref{fig:walsh_cshunt_coupler} shows the interaction coefficients $w_{\bm{n}}$ for the qubit ladder with model parameters determined in this manner. As expected, the overall reduction of $ZZ$ interactions is significantly stronger than for tunable transmon couplers, with the nearest-neighbor terms remaining below $20\,$kHz for $T \leq 20\,$MHz. This corresponds to a reduction of at least one order of magnitude compared to conventional tunable couplers. 
The specific long-ranged terms discussed  
  for the models with static coupling and tunable transmon couplers are also strongly suppressed, which is perhaps 
  related to an enhanced localization of the dressed qubits due to the minimization of the delocalization parameter $\epsilon$ in Eq.~\eqref{eq:delocalization_parameter}. 
  However, the IPR does not differ much compared to the model with transmon couplers (Fig.~\ref{fig:IPR}), likely because our parameter choice results in very similar hopping amplitudes for the bosons in both models, and is in fact slightly lower for $T \lesssim 12$\,MHz. 
  The mutual information, on the other hand, is noticeably smaller, especially at larger $T$. 
  Together with the results for the interaction coefficients $w_{\bm{n}}$, this suggests that the C-shunt flux couplers are significantly more effective in reducing the buildup of longer-ranged correlations with $T$ than transmon couplers (whose parameters are optimized to minimize the $ZZ$ interaction). To fully confirm that this is a general difference between the two coupling schemes, one would have to check additional frequency patterns, however.

\begin{figure}[tb]
  \centering
  \includegraphics[width=0.97\linewidth]{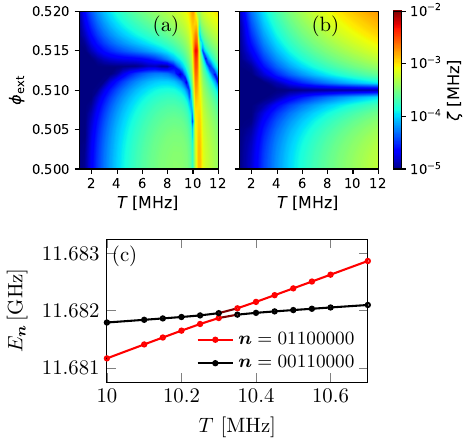}
	\caption{ 
	(a) Approximate $ZZ$ coupling $\zeta$ defined in Eq.~\eqref{eq:poormanswalsh} between qubits $3$ and $4$ as a function of external flux and capacitive coupling strength. (b) $ZZ$ coupling for the isolated qubit pair. 
	The avoided level crossing responsible for the jump in $\zeta$ is shown in panel (c).
        } 
  \label{fig:bond2-3}
\end{figure}

  Since there is still a systematic increase of the residual $ZZ$ couplings with $T$, 
  an obvious question is whether a further reduction is possible by adjusting the coupler parameters using the information of the full system. Following Ref.~\cite{CShuntFluxCoupler}, where such a fine-tuning was demonstrated for a short chain of 4 qubits, we take the fluxes $\phi_{\text{ext}}$ as controllable parameters. 
To keep the numerical optimization for the full system manageable, we do not calculate the entire computational basis and instead define a cost for each coupler as
\begin{align}
 \zeta = \frac{1}{4} \big| E_{11}' - E_{10}' - E_{01}' + E_{00}' \big| \, ,
 \label{eq:poormanswalsh}
\end{align}
where $E_{n_1 n_2}'$ is the energy of the state, in which the qubits connected to the coupler are in the states $n_1$ and $n_2$, and all other qubits are in the $0$ state. This becomes equal to the absolute values of nearest-neighbor couplings in Eq.~\eqref{eq:l-bits} when higher-order terms involving 3 or more qubits vanish. 
As an example, Fig.~\ref{fig:bond2-3}(a) displays $\zeta$ as a function of $\phi_{\text{ext}}$ and $T$ for the qubit pair $2$ and $3$. For small $T$, the difference compared to the isolated qubit dimer [Fig.~\ref{fig:bond2-3}(b)] is mostly a change of the optimal value for $\phi_{\text{ext}}$. Around $T\approx 10$\,MHz, however, there is a sudden increase of $\zeta$ caused by 
an avoided crossing between the eigenstates for $\bm{n}=01100000$ and $\bm{n}=00110000$ [Fig.~\ref{fig:bond2-3}(c)], showing that the effect of spectator qubits cannot always be accounted for by a simple shift of $\phi_{\text{ext}}$. 
It should also be noted that the optimization only addresses the nearest-neighbor $ZZ$ interactions, so we can not reasonably expect it to remove all stray couplings for larger $T$, where longer-ranged and multi-qubit terms become important.  
Nevertheless, the DMRG simulations confirmed that minimizing $\zeta$ independently for each coupler while keeping the remaining model parameters fixed is sufficient to bring all interaction strengths $|w_{\bm{n}}|$ below 50\,Hz for weak capacitive coupling $T \leq 8$\,MHz. 

\section{Fluxonium with transmon couplers}
\label{sec:fluxonium}
Fluxonium is a different type of superconducting qubit, in which a Josephson junction with energy $E_J$ is shunted with an array of larger junctions that act as an inductance $E_L$~\cite{Fluxonium}. 
A single fluxonium is described by the Hamiltonian 
\begin{align}
	\hat{H}_{\text{F}} &= 4 E_C \hat{n}^2 - E_J \cos(\hat{\phi} + 2 \pi \Phi_\text{ext} / \Phi_0 ) + \frac{1}{2} E_L \hat{\phi}^2 \, ,
	\label{eq:fluxonium}
\end{align}
where $\Phi_\text{ext}$ is the external flux threading the qubit loop. 
We assume $\Phi_\text{ext} = \Phi_0 /2$, so that the cosine term becomes a double-well potential for the flux variable. 
In numerical simulations, we apply a truncation for each fluxonium in the eigenbasis of its single-qubit Hamiltonian~\eqref{eq:fluxonium}, keeping the $8$ lowest eigenstates.

Potential advantages fluxonium qubits offer over transmons are long coherence times~\cite{PhysRevX.9.041041}, and a higher nonlinearity on the order of a few GHz, which helps preventing leakage into higher energy levels outside the computational subspace. 
While larger fluxonium devices have yet to be realized, several different schemes have been proposed to implement two-qubit gates and were demonstrated for systems of two qubits~\cite{FluxoniumTunableCoupler,FluxoniumBlueprint,FluxoniumWithTransmonCoupler,FluxoniumTunableInductiveCoupler}. 
Recently, it was shown that high-fidelity controlled-Z (CZ) gates can be implemented in an architecture with flux-tunable transmons as couplers~\cite{FluxoniumWithTransmonCoupler}. 
An attractive feature of this architecture is the suppression of the static $ZZ$ interaction, which is otherwise strong 
for capacitively coupled fluxonium qubits because of the large matrix elements of the charge operator $\hat{n}$ between computational and non-computational states~\cite{FluxoniumBlueprint}.  
To investigate how robust the reduction of the $ZZ$ coupling is when going to larger systems, 
we simulate a device of this type with the same ladder geometry as in the previous section (see Fig.~\ref{fig:sys2x4}). 
Its Hamiltonian has the form~\eqref{eq:qubitlattice}, with the local terms given by Eq.~\eqref{eq:fluxonium} and Eq.~\eqref{eq:duffing} for the fluxonium qubits and transmon couplers, respectively.

\begin{table}[tb]
  \centering
	\newcommand{\taborder}[6]{ #1 & #2 & #4 & #3 & #5 & #6}
  \setlength{\tabcolsep}{8pt}
  \begin{tabular}{c|ccccc}
	   \taborder{qubit}{$E_C$}{$E_L$}{$E_J$}{$\nu$}{$\alpha$}\\
    	  \hline
	  \taborder{$q_1$}{1.05}{0.93}{5.54}{0.254}{4.270} \\ 
	  \taborder{$q_2$}{ 1.18}{ 1.35}{ 5.36}{ 0.635}{ 3.649} \\ 
	  \taborder{$q_3$}{ 0.92}{ 0.98}{ 4.57}{ 0.352}{ 3.272} \\ 
	  \taborder{$q_4$}{ 0.97}{ 1.31}{ 4.97}{ 0.512}{ 3.231} \\ 
	  \taborder{$q_5$}{ 1.05}{ 1.21}{ 5.02}{ 0.505}{ 3.449} \\ 
	  \taborder{$q_6$}{ 0.86}{ 0.94}{ 4.48}{ 0.302}{ 3.207} \\ 
	  \taborder{$q_7$}{ 1.14}{ 1.29}{ 5.29}{ 0.573}{ 3.638} \\ 
	  \taborder{$q_8$}{ 0.82}{ 0.79}{ 4.11}{ 0.261}{ 3.059} \\[1mm]
	  \taborder{$c_{12}$}{ 0.33}{ - }{ 17.8}{6.525}{-0.33}\\
	  \taborder{$c_{15}$}{ 0.28}{ - }{ 17.7}{6.017}{-0.28}\\
	  \taborder{$c_{26}$}{ 0.32}{ - }{ 18.8}{6.617}{-0.32}\\
	  \taborder{$c_{56}$}{ 0.29}{ - }{ 17.6}{6.100}{-0.29}\\
	  \taborder{$c_{23}$}{ 0.27}{ - }{ 19.2}{6.170}{-0.27}\\
	  \taborder{$c_{37}$}{ 0.28}{ - }{ 19.7}{6.363}{-0.28}\\
	  \taborder{$c_{67}$}{ 0.31}{ - }{ 15.1}{5.809}{-0.31}\\
	  \taborder{$c_{34}$}{ 0.29}{ - }{ 20.3}{6.573}{-0.29}\\
	  \taborder{$c_{48}$}{ 0.31}{ - }{ 15.5}{5.890}{-0.31}\\
	  \taborder{$c_{78}$}{ 0.29}{ - }{ 17.3}{6.045}{-0.29}
  \end{tabular}
	\caption{Parameters used for the system with fluxonium qubits $q_i$ and transmon couplers $c_{ij}$ (in GHz). How the qubit labels map to the ladder is shown in Fig.~\ref{fig:sys2x4}. }
  \label{table:fluxoniumparameters}
\end{table}

\begin{figure}[tb]
  \centering
  \includegraphics[width=0.92\linewidth]{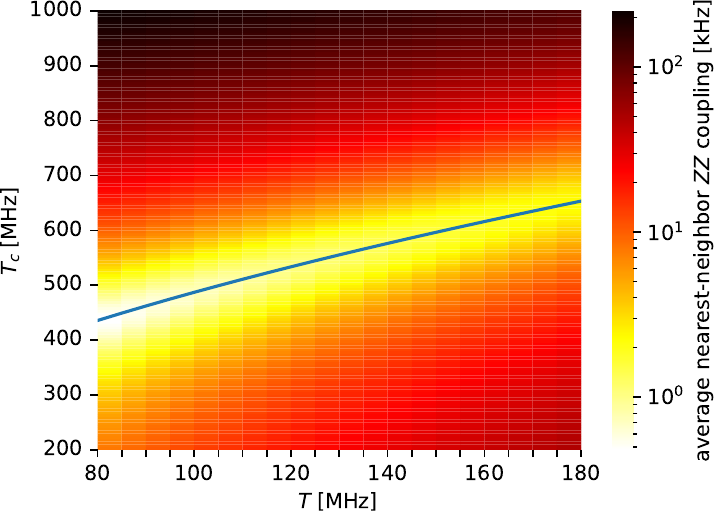}
  \caption{$ZZ$ coupling averaged over decoupled subsystems of nearest-neighbor fluxonium qubits and the connecting transmons as a function of the coupling strengths $T$ and $T_c$. The line shows the $T$-$T_c$ curve used for the MPS simulations of the full system.}
  \label{fig:fluxonium_parameter_scan}
\end{figure}

When balancing the parameters of fluxonium qubits, several aspects must be considered. A sufficiently small $E_C$ allows for larger coupling capacitances, while $E_L$ should not be chosen too large in order to avoid smaller coherence times. In this specific example, we employ a quasi-periodic arrangement of qubit frequencies to mitigate the risk of frequency crowding. We emphasize, however, that there exists a broad parameter regime, and the precise choice may ultimately depend on the requirements of a specific gate scheme. Based on these considerations we propose the set of parameters listed in Table~\ref{table:fluxoniumparameters}. 
The strengths of the capacitive couplings are again uniform. 
For each $T$, we first determine the value of $T_c$ that minimizes the average $ZZ$ coupling between nearest-neighbor qubits 
when treating the qubit dimers as isolated (see Fig.~\ref{fig:fluxonium_parameter_scan}). 
We then use the parameters in Table~\ref{table:fluxoniumparameters} along with the optimized coupling $T_c$ for the MPS simulation of the full system. 

\begin{figure}[tb]
  \centering
  \includegraphics[width=0.92\linewidth]{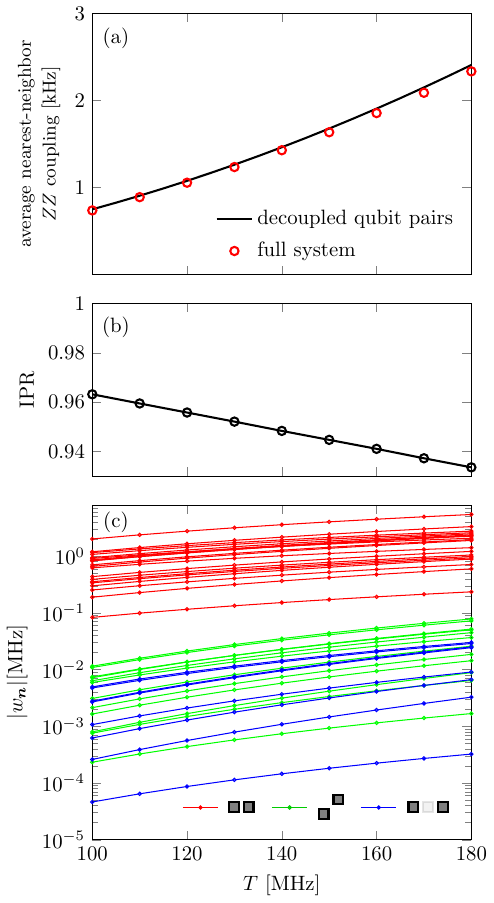}
  \caption{Average strength of the $ZZ$ interactions between nearest-neighbor fluxonium qubits in the ladder system compared with the corresponding interactions for isolated qubit pairs (a), and average IPR of the computational basis states for the same system (b). Panel (c) shows couplings involving higher states as described in the main text. }
  \label{fig:fluxonium_ladder}
\end{figure}

Figure~\ref{fig:fluxonium_ladder} shows the obtained nearest-neighbor $ZZ$ couplings and the average IPR. 
The $ZZ$ couplings are on the order of a few kHz and deviate only slightly from the results for the isolated qubit dimers. 
Moreover, the additional longer-ranged interactions and those involving three or more qubits have coefficients $w_{\bm{n}}$ with magnitude below 10\,Hz, and are thus negligible. 
The IPR remains at a large value $> 0.9$ for all considered parameters, indicating that the computational basis states 
remain strongly localized. 
This can likely be attributed to the fact that the capacitive coupling only weakly mixes the $|0\rangle$ and $|1\rangle$ states of the fluxonium, and that the additional transmons have a much larger transition frequency, so that the hybridization with the couplers is also suppressed. 

While the results are promising regarding the scalability of a fluxonium-based system, a direct comparison with the transmon systems studied in Sec.~\ref{sec:transmons} is difficult because of the specific parameter choices involved in our simulations. Furthermore, the early stage of fluxonium research leaves several crucial questions unanswered, and unlike for transmon systems, a large-scale fluxonium chip has not yet been demonstrated. 
Key challenges include tackling the increased risk of frequency crowding and finding the best approach for realizing two-qubit gates. Reference~\cite{FluxoniumWithTransmonCoupler} proposes a CZ gate implementation, in which the $|11\rangle$ state of a fluxonium pair is selectively driven to a state outside the computational subspace and back. This approach leverages the fact that couplings to the second-excited subspace are significantly stronger in fluxonium systems compared to the bosonic type interactions in transmon systems. 
However, this method introduces two scaling challenges: ensuring that these higher-energy states remain addressable in a larger-scale chip and resolving the capacitance budgeting issue commonly associated with fluxonium architectures. 
Although these questions fall outside the scope of this work, we emphasize that states in the two-excitation subspace tend to be in a more delocalized regime, which may have important implications for scalability.

Taking the corresponding eigenstates of the full system into account
one can define an effective Hamiltonian similar to~\eqref{eq:l-bits} by including terms with operators $\hat{P}_i = \sum_{\bm{n} \in \{0,1,2\}^N} \delta_{n_i2} |\tilde{\bm{n}}\rangle \langle \tilde{\bm{n}} |$. 
As an example, Fig.~\ref{fig:fluxonium_ladder}(c) shows the coefficients for interactions of the form ${\propto} \hat{Z}_i \hat{P}_j$, which would be relevant when implementing the CZ gate mentioned above. 
The calculation only requires eigenstates with at most one qubit in the $|2\rangle$ state, significantly reducing the number of DMRG simulations. 
For our model, we find that the nearest-neighbor couplings are on the order of $1\,$ MHz, while the next-nearest neighbor ones reach values around a few tens of kHz. These stronger interactions are expected because of the reduced localization of the states and the fact that the parameters were chosen to minimize the $ZZ$ coupling specifically.

\section{Discussion}
\label{sec:discussion}
Building a superconducting quantum computer requires achieving fast gate operations while minimizing unwanted crosstalk across the qubit chip. Although increasing coupling strengths enables faster gates, it also amplifies crosstalk and may even induce chaotic instabilities in larger systems, making it challenging to strike an optimal balance. 
For fluxonium qubits, we showed that a transmon-based tunable coupler can suppress $ZZ$ crosstalk to the level of $\sim1\,$kHz. However, to contextualize this result, particularly in comparison to transmon architectures, we must assess the coupling strengths required for implementing gates with similar speeds. In Ref.~\cite{FluxoniumWithTransmonCoupler}, a coupling strength of $T \approx 120\,$MHz was used to implement a 60\,ns CZ gate for two Fluxoniums. In contrast, a similar CZ gate in a transmon system with transmon tunable couplers, as demonstrated in Ref.~\cite{PhysRevLett.125.240502}, was achieved in a 38\,ns gate time with $T \approx 20\,$MHz. Notice that both gates were studied in isolated two-qubit systems. Our simulations suggest that in a scaled-up qubit layout, residual crosstalk in a fluxonium system would be an order of magnitude lower, indicating that static $ZZ$-crosstalk might not be the primary bottleneck for scaling fluxonium-based architectures. 
For a more comprehensive study of the scalability perspectives of fluxonium architectures, it is of course important to also investigate other sources of cross-talk, e.g. between the dc flux lines for controlling the qubits and couplers. 
It should also be noted that by treating the nearest-neighbor coupling as an independent parameter, we did not account for capacitive loading~\cite{FluxoniumResonatorCoupler}, which can become a problem when scaling up the number of qubits as it limits the attainable coupling strengths for fluxoniums with higher connectivity. 

In our simulations, we neglected the longer-ranged terms in the Hamiltonian that should become more relevant in both transmon and fluxonium systems as the coupling capacitances increase. 
As an example, consider the transmon architecture with tunable couplers at capacitive coupling strength $T=10\,$MHz. 
Inverting the capacitance matrix that (approximately) corresponds to this model yields next-nearest-neighbor transverse couplings below $100\,$kHz. 
In perturbation theory, this causes corrections to the next-nearest-neighbor $ZZ$ interactions $w_{\bm{n}}$ in the range 10\,Hz--100\,Hz, which is on the order of those coefficients $w_{\bm{n}}$ in the model with only nearest-neighbor couplings, but well below the nearest-neighbor $ZZ$ interactions. 
The longer-ranged couplings therefore do not appear to significantly affect our results.  
However, because of frequency collisions, perturbation theory might underestimate the corrections for some qubit pairs. Here, further numerical analysis would yield more accurate quantitative predictions for these corrections.

Lastly, let us comment on the geometry of the system. 
Although the $2\times 4$-ladder considered in this work is small compared to an actual large-scale device, it already has an important feature of two-dimensional systems in that it includes loops with multiple coupling paths between qubits. 
In the regime we are interested in, where the qubits are localized but first signs of delocalization appear, we therefore expect the ladder model to give meaningful results for the $ZZ$ crosstalk in qubit lattices. 
The larger connectivity in a two-dimensional system may affect localization measures like the IPR, but since we always used the same geometry, the comparisons between architectures should still be valid.

\section{Summary}
\label{sec:summary}
We used density-matrix renormalization-group simulations to investigate how well different schemes for superconducting qubit devices are able to suppress the static $ZZ$ interaction when the system is scaled up. 
In agreement with previous works, we found that lattices of fixed-frequency transmons are limited to small capacitive couplings on the order of a few MHz, before the nearest-neighbor $ZZ$ interaction becomes prohibitively large. Although the addition of tunable couplers can reduce the effective interactions between qubits significantly, the qubits will still delocalize at sufficiently large capacitive couplings, and we indeed observe emerging long-range interactions due to near-resonant next-nearest neighbor qubits. Avoiding such frequency collisions should therefore also be beneficial for architectures with tunable couplers. 

We considered parameters outside the straddling regime with relatively large detuning between nearest-neighbor qubits. Choosing the qubit frequencies in this manner helps keeping the computational basis states localized but also prevents the tunable transmon couplers from completely suppressing the static interactions already for a single qubit pair. 
Using C-shunt flux qubits as tunable couplers leads to much smaller residual couplings, while requiring more complex circuits. 

For the architecture with fluxonium qubits and tunable transmon couplers, the results are promising, showing an efficient suppression of the $ZZ$ interaction and a large IPR over a wide range of capacitive coupling strengths. 

As direction for future study, it would be interesting to investigate how longer-ranged couplings influence the delocation as coupling strengths are increased. In principle, such terms could be easily included in the DMRG simulations. 
It would also be possible to extend the MPS approach used here to time-dependent simulations~\cite{TDVPMPS}, e.g., 
to investigate the effect of stray couplings during gate operations~\cite{PhysRevApplied.19.024057}.

\begin{acknowledgments}
This work has received
funding from the German Federal Ministry of Education and Research via the funding program quantum
technologies - from basic research to the market under
contract numbers 13N15684 ”GeQCoS” and 13N16182
”MUNIQC-SC”. It is also part of the Munich Quantum Valley, which is supported by the Bavarian state government with funds from the Hightech Agenda Bayern Plus. 
The authors gratefully acknowledge the scientific support and HPC resources provided by the Erlangen National High
Performance Computing Center (NHR@FAU) of the Friedrich-Alexander-Universit\"at Erlangen-N\"urnberg (FAU). NHR funding is provided by federal and Bavarian state authorities. NHR@FAU hardware is partially funded by the German Research Foundation (DFG) – 440719683.   
MPS simulations were performed using the ITensor library~\cite{itensor,itensor-r0.3}. 
\end{acknowledgments}

\section*{Data availability statement}
All data that support the findings of this study are included within the article.

\end{document}